\documentclass[authoryear,preprint,review,2.5p,12pt]{elsarticle} 
\usepackage[a4paper, margin=1in, left=1in, right=1in]{geometry}

\usepackage[utf8]{inputenc}
\usepackage{xcolor}
\usepackage{float}
\usepackage{mathtools}
\usepackage{subcaption}
\usepackage{todonotes}
\usepackage{psfrag}
\usepackage{ulem}
\usepackage{gensymb}
\usepackage{amsmath}

\setcitestyle{square,numbers,sort&compress,comma}
\newcommand{\str}[1]{}

\newcommand{\makered}[1]{\textcolor{red}{#1}} 
\usepackage{comment}
\usepackage{wasysym}

\date{June 2024}

\newcommand{\OHstar}{$\mathrm{OH^{*}}$}
\newcommand{\YHsecond}{Y_{\mathrm{H_2}}^{\mathrm{2nd}}}
\newcommand{\YHfirst}{Y_{\mathrm{H_2}}^{\mathrm{1st}}}

\begin{document}
\begin{frontmatter}
\title{Flame Transfer Function Measurement of a Hydrogen-Blended Sequential Combustor}
\author{Bayu Dharmaputra\corref{cor1}}
\ead{bayud@ethz.ch}
\author{Pushkin Nagpure\corref{}}
\author{Matteo Impagnatiello\corref{}}

\author{Nicolas Noiray\corref{cor1}}
\ead{noirayn@ethz.ch}
\cortext[cor1]{Corresponding authors}

\address{CAPS Laboratory, Department of Mechanical and Process Engineering, ETH Z\unexpanded{\"u}rich, 8092, Z\unexpanded{\"u}rich, Switzerland}
\begin{abstract}

The flame transfer function (FTF) relates acoustic perturbations and the coherent heat release rate response. This frequency-dependent function governs the thermoacoustic stability of a combustor. The FTF measurement is therefore of great interest for predicting the stability of the a practical combustor connected to the engine's compressor and turbine. So far, the FTFs of the second stage of constant pressure sequential combustors (CPSC) have only been obtained from numerical simulations. In this study, second-stage FTFs are measured experimentally. The thermal power of both the first- and second-stage flames is kept constant and their ratio is fixed at 1:1. The effects of hydrogen blending in the first and second stage fuel mixture on the sequential FTF are analyzed. The FTF of the sequential flame is fitted with a distributed time delay (DTD) model with two pulses. The trends of the model parameters obtained are consistent with the chemiluminescence \OHstar{} of the sequential flame.

\end{abstract}

\end{frontmatter}
\newpageafter{abstract}

\section*{Novelty and significance}

In this study, the flame transfer function of the second-stage flame of an atmospheric sequential combustor is measured for the first time. This is a significant progress as the second stage FTF had been obtained so far only from numerical simulations. The effects of the amount of hydrogen mixture in the first- and second-stage fuels on the sequential flame transfer function are investigated. This work thus serves as a stepping stone towards the thermoacoustic modeling of a sequential combustor for stability prediction.

\section*{Authors contributions}

B.D and N.N. conceived the research idea. B.D. led the experimental investigations. B.D. P.N. and M.I performed the experiments. B.D and P.N analyzed the data. B.D., P.N., M.I. and N.N. discussed the results. B.D. drafted the manuscript with the support of N.N. The final version of the manuscript has been edited and approved by B.D., P.N., M.I. and N.N.

\section{Introduction}

Modern gas turbines for electric power production must achieve fast ramp-up and ramp-down times to balance the inherent intermittency of wind and solar power sources. The constant pressure sequential combustor (CPSC) is a promising solution to these challenges \cite{Pennell2017}. CPSC offers numerous advantages, including high fuel flexibility, low NOx and CO emissions over a wide operating range \cite{Ciani2019, Bothien2019}.

However, as in single-stage combustor architectures, the operational range of CPSCs is limited by thermoacoustic instabilities. One of the key information that needs to be quantified to predict these thermoacoustic instabilities is the flame transfer function (FTF). The FTF relates the heat release rate fluctuations of the flame to the upstream velocity perturbations. The FTFs of single-stage combustors have been widely studied in both laboratory-scale setups, e.g. \cite{Blonde2023,DISABATINO2018,MOON2024,zurNedden2023,AESOY2022} and industrial setups \cite{Schuermans2010,Tony2023}.

To the best of our knowledge, the FTF of CPSC's second stages have only been characterized by numerical simulations \cite{Yang2015,Gant2022,IMPAGNATIELLO2024,SCHULZ2018} and analytical methods \cite{GOPALAKRISHNAN2023,GANT2021}, and no experimental FTFs have been measured so far.

In this study, we aim at filling this knowledge gap by experimentally measuring the second-stage FTF of an atmospheric sequential combustor. In particular, we investigate the effect of hydrogen blending in both first- and second-stage fuel lines on the FTF. Furthermore, we fit the measured FTFs using the widely known distributed time delay (DTD) model \cite{POLIFKE2020}, and show that for the forcing amplitudes considered here, the response of the reheat flame to acoustic excitation exhibits the characteristics of a technically premixed first-stage flame with stiff fuel injection.

\section{Experimental setup}
\begin{figure}[t!]
    \centering
    \def\svgwidth{1.1\textwidth}
    \input{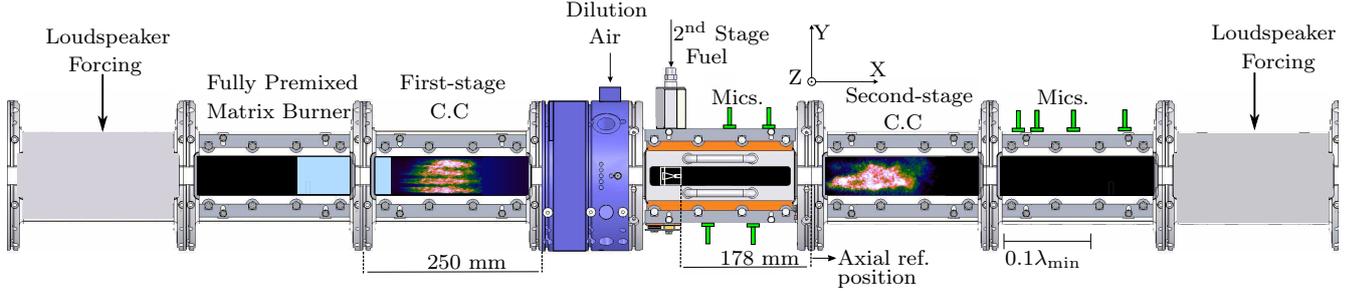}
    \caption{Experimental setup for sequential flame transfer function investigations. C.C: combustion chamber, Mics.: Microphones, $\lambda_{\mathrm{min}} = c_{ds}/f_{\mathrm{max}}$.}
    \label{fig:exp_setup}
\end{figure}
The sequential combustor setup, similar to that in \cite{DHARMAPUTRA2023,DHARMAPUTRA2024}, is briefly described here. Figure~\ref{fig:exp_setup} shows the side view of the setup for FTF measurements. It includes a plenum with loudspeakers for acoustic forcing, a fully premixed burner composed of a 4$\times$4 matrix of $\diameter$6~mm, a combustion chamber with a 62×62 mm² cross-section, and a sequential burner with a mixing channel (25×38 mm² cross-section). Secondary fuel is injected 178~mm from the burner outlet, through an X-lobe injector with a diameter of 1.5~mm creating a stiff injection. A second pair of loudspeakers is located downstream of the second-stage combustion chamber. Both sets of loudspeakers provide an acoustic forcing from 190~Hz to 700~Hz with a step of 10~Hz. First-stage air with a mass-flow of 17~g/s is preheated to 250°C, and a hydrogen-natural gas blend is injected upstream of the loudspeakers, creating a premixed mixture. Dilution air with a mass flow of 18~g/s is injected downstream of the first-stage combustion chamber. 

The tunable diode laser absorption spectroscopy with wavelength modulation spectroscopy (TDLAS-WMS) method, which can measure relatively low coherent temperature fluctuations down to 5~K as detailed in \cite{DHARMAPUTRAPROCIETF}, measures temperature fluctuations 10~cm downstream of the second-stage fuel injection in the mixing channel, aligned similarly to \cite{SOLANAPEREZ2022}. However, no coherent temperature fluctuations are detected in the acoustic frequency band of interest and for the acoustic forcing amplitudes considered. 

Four flush-mounted G.R.A.S microphones are placed along the sequential mixing channel walls, with another four downstream of the second-stage combustion chamber. Both sets of microphones are used to reconstruct the acoustic variables upstream and downstream of the reference location, which in this case is the sequential burner outlet, with the multi-microphone method (MMM) \cite{Paschereit2002}. A high-speed intensified camera is pointed at the sequential flame to capture the \OHstar~chemiluminescence.

The hydrogen blending mass fractions for the first and second stage fuel, denoted by $\YHfirst$ and $\YHsecond$, are defined as the ratio of hydrogen mass flow to the total mass flow of hydrogen and natural gas in each fuel line. In this study, both quantities are varied, but the thermal power of both flames is kept constant at 77~kW and in a 1:1 ratio for all operating conditions. Both flames are in lean condition, with an equivalence ratio range of 0.74-0.77 and 0.54-0.6, for the first flame and sequential flame, respectively. The flow velocity inside the mixing channel remains around 110 m/s and with a temperature of about 1200~K for all cases. It is worth mentioning that for all operating conditions, no ignition kernel is generated inside the mixing channel.

\section{Methodology}

 The relative heat release rate fluctuations, $\dot{Q}'/\dot{\overline{Q}}$, can be written as a function of the relative upstream pressure ($p_{us}'/\overline{p}_{us}$), velocity ($u_{us}'/\overline{u}_{us}$), equivalence ratio ($\phi'/\overline{\phi}$), and temperature fluctuations ($T_{us}'/\overline{T}_{us}$) as discussed in \cite{Gant2022,SCHULZ2018, Bothien2018}. We adopt here the description used in \cite{Bothien2018}. It does not explicitly account for equivalence ratio fluctuations, whose effects are embedded in the other transfer functions, as will be shown later.
\begin{equation}
\frac{\dot{Q}'}{\dot{Q}} = F_p \frac{p_{us}'}{\overline{p}_{us}} + F_u \frac{u_{us}'}{\overline{u}_{us}} + F_T \frac{T_{us}'}{\overline{T}_{us}}.
\end{equation}

As mentioned previously, the TDLAS-WMS does not detect coherent temperature fluctuations, $T'_{us}$, in the frequency and forcing amplitude ranges of interest. Therefore, in this study, $F_T$ is not quantified and we could resort to the classical 2$\times$2 flame transfer matrix (FTM) identification approach as described in \cite{Blonde2023,MOON2024}. From the Rankine Hugoniot relationships, the acoustic velocity fluctuations downstream of the sequential flame ($u_{ds}'$) can be written as a function of the $F_{21}$ and $F_{22}$ elements of the flame transfer matrix (FTM):

\begin{equation} \label{eq:FTM_2}
\begin{split}
        u_{ds}' &= F_{21}\frac{p'_{us}}{(\bar{\rho}\bar{c})_{us}} + F_{22} u_{us}'\\
            &= \gamma\theta M_{us} (F_p-1)\frac{p'_{us}}{(\bar{\rho}\bar{c})_{us}} + (1+\theta F_u) u_{us}'\\
        \theta  &\coloneqq \frac{\overline{T}_{ds}}{\overline{T}_{us}} -1,
\end{split}
\end{equation}
where $\gamma$ and $M_{us}$ is the ratio of the specific heat capacities, and the upstream Mach number, respectively. 

The $F_{21}$ element of the FTM that we measured exhibits a very low gain of about 5\% of $|F_{22}|$ in the frequency range of interest for all operating conditions. Hence, in this study, we focus only on the $F_{22}$ element. It is worth mentioning that in our measurement, the relative upstream velocity perturbations, $u_{us}'/\overline{u}_{us}$, range around 1-3\% for the frequency ranges and all operating conditions considered. The $F_u$ in Eq.~(\ref{eq:FTM_2}) is the classical FTF and can be obtained from $F_{22}$: $F_u = (F_{22}-1)/\theta$. The obtained FTF is then fitted to a DTD model, described in \cite{POLIFKE2020}, with two Gaussian pulses:

\begin{equation}\label{eq:DTD_model}
    \mathrm{FTF}(\omega) = n_1 \exp\bigg(-i\omega\tau_1-\frac{\omega^2\sigma_1^2}{2}\bigg) + n_2 \exp\bigg(-i\omega\tau_2-\frac{\omega^2\sigma_2^2}{2}\bigg),
\end{equation}
where $\tau_k$, and $\sigma_k$ model the time delay and the characteristic spread of each pulse, respectively. By looking again at the experimental setup in figure~\ref{fig:exp_setup}, one could see that the sequential flame is close to a technically premixed configuration . This implies that the upstream acoustic perturbations would additionally generate equivalence ratio perturbations $\phi'$. Therefore, FTF gain at 0~Hz must equal to zero under the assumption of a stiff second-stage fuel injection \cite{POLIFKE2007}. An example of FTF obtained through numerical simulation that is similar to our configuration could be seen in \cite{Yang2015}, and in that study the gain of FTF at 0~Hz is zero. In order to fulfill this criteria, a constraint is added to the FTF model:
\begin{equation}\label{eq:n_constraint}
    n_1+n_2 = 0, \mathrm{and}~n_1 \equiv n
\end{equation}

To characterize the flame length and spread, the 2D image of the \OHstar{} chemiluminescence, $I (x,y)$, is integrated along the vertical axis (Y-axis in figure~\ref{fig:exp_setup}): I(x) = $\int_{-H/2}^{H/2} I(x,y) dy$. The center of mass of the flame, $\mu_f$, and the characteristic width, $\sigma_f$, are then computed as: 

\begin{align}
    \mu_f &= \frac{\int x I(x)dx}{\int I(x) dx}\\
    \sigma_f^2 &= \frac{\int (x-\mu_f)^2 I(x)dx}{\int I(x) dx}.
\end{align}

\section{Results and Discussion}

\subsection{Variations of hydrogen blending in the sequential fuel line}

\begin{figure}[H]
    \centering
    \def\svgwidth{0.5\textwidth}
    \input{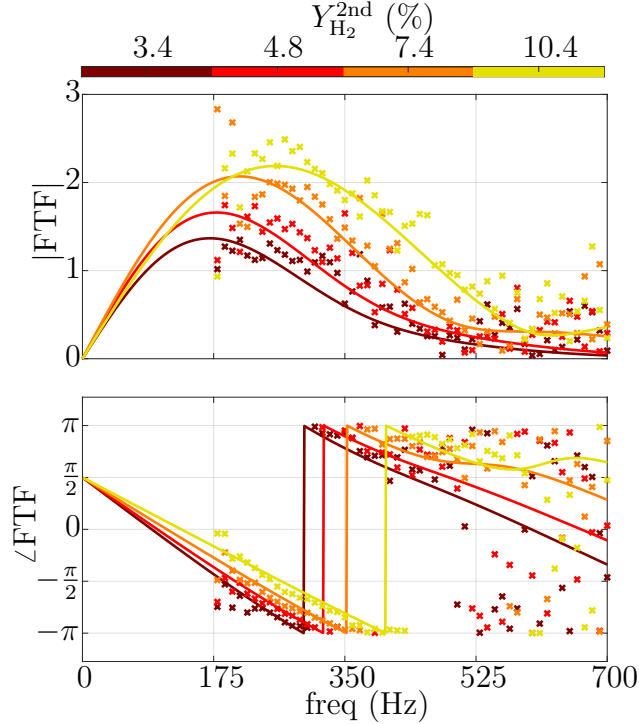}
    \caption{Measured (symbols) and fitted FTFs (solid lines) for different $\YHsecond$. The $\YHfirst$ is fixed at 4.8\%. Two DTDs are employed for the fitting according to Eqs.~(\ref{eq:DTD_model}) and~(\ref{eq:n_constraint}). The gain and phase of FTFs are shown on the top and bottom plot, respectively.}
    \label{fig:FIT_FTF_YH2_2ndstage}
\end{figure}

The measured FTFs and their corresponding best fit with Eq.~(\ref{eq:DTD_model}) for different $\YHsecond$ are shown in Figure~\ref{fig:FIT_FTF_YH2_2ndstage}. Note that in these cases, the $\YHfirst$ is fixed at 4.8\%. As can be seen, the gain of FTF exhibits noticeable variations with respect to the variations in $\YHsecond$. Furthermore, the slope of the phase becomes flatter as $\YHsecond$ increases. This indicates that there is a decrease in the characteristic time delay of the perturbations, which can be related to the change in the shape sequential flame as will be discussed below.

Figure~\ref{fig:combined_figures_2nd_var}a shows the DTD parameters of the FTF fitting. Figure~\ref{fig:combined_figures_2nd_var}b shows the vertically integrated flame chemiluminescence at different hydrogen fractions in the sequential fuel line. As can be seen, the flame moves closer to the burner outlet as $\YHsecond$ increases. This behavior is expected as the fuel reactivity also increases. Both the center of mass and the spread of the flame exhibit monotonically decreasing trends with respect to $\YHsecond$.
The trends of $\sigma$ and $\tau$ from the fitting of the model are consistent with that of the flame chemiluminescence as can be deduced by looking at figure~\ref{fig:combined_figures_2nd_var}a~and~\ref{fig:combined_figures_2nd_var}b. 

\begin{figure}[H]
    \centering
    \vspace{-0.5cm}
    \begin{subfigure}{0.4\textwidth}
        \centering
        \def\svgwidth{1\textwidth}
        \input{param_opt_YH2_2ndStage_var_}
    \end{subfigure}%
    \hfill
    \begin{subfigure}{0.5\textwidth}
        \centering
        \def\svgwidth{0.94\textwidth}
        \input{COM_2ndstage_var_}
    \end{subfigure}
    \caption{(a) The resulting mean time delay $\tau$, time delay spread $\sigma$'s and gain $n$ parameters from the fitting of the flame transfer function across different hydrogen blending in the first-stage fuel line, $\YHsecond$. (b) The vertically integrated flame chemiluminescence at different $\YHsecond$ (top) along with the flame centre of mass and characteristic width (bottom).}
    \vspace{-0.5cm}
    \label{fig:combined_figures_2nd_var}
\end{figure}

Indeed, the mean time delay parameters $\tau_1$ and $\tau_2$ decrease with increasing hydrogen fraction in the sequential fuel line, attributed to flame shortening as shown in Figure~\ref{fig:combined_figures_2nd_var}b. The difference between the time delays, $\Delta\tau = \tau_2 - \tau_1$, remains constant at approximately 1.6~ms across different hydrogen blending amounts. This matches the expected convective delay, $\tau_{conv} =L/\overline{U}_{conv}~= 0.178/110 \approx 1.6$~ms, given the flow velocity of 110~m/s and the 178~mm distance from the sequential fuel injector to the dump plane.

\subsection{Variations of hydrogen blending in the first stage fuel line.}

\begin{figure}[H]
    \centering
    \vspace{-1cm}
    \def\svgwidth{0.46\textwidth}
    \input{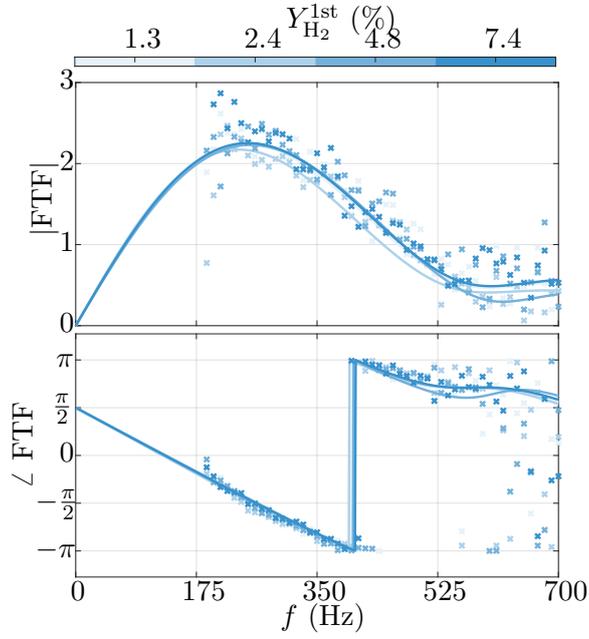}
    \caption{Measured (symbols) and fitted FTFs (solid lines) at different $\YHfirst$. $\YHsecond$ is fixed at 10.45\%. Two DTDs are employed for the fitting. The gain and phase of FTFs are shown on the top and bottom plot, respectively.}
    \label{fig:FIT_FTF_YH2_1ststage}
    \vspace{-0.4cm}
\end{figure}

In the second test case, $\YHsecond$ is fixed at 10.45\%, and $\YHfirst$ is varied from 1.3 to 7.4\%. Note that the thermal power of the first stage flame is kept constant, and hence, the temperature in the mixing channel is only weakly affected. The measured FTFs and the resulting model fittings can be seen in figure~\ref{fig:FIT_FTF_YH2_1ststage}. In contrast to the previous case, both the gain and phase of the FTFs show no significant variation with respect to changes in $\YHfirst$. 

\begin{figure}[H]
    \centering
    \begin{subfigure}{0.4\textwidth}
        \centering
        \def\svgwidth{0.95\textwidth}
        \input{param_opt_YH2_1stStage_var_}
    \end{subfigure}%
    \hfill
    \begin{subfigure}{0.5\textwidth}
        \centering
        \def\svgwidth{0.95\textwidth}
        \input{COM_1ststage_var_}
    \end{subfigure}
    \caption{(a) The resulting time delay $\tau$'s spread $\sigma$'s and gain $n$ parameters from the fitting of the flame transfer function across different hydrogen blending in the first-stage fuel line, $\YHfirst$. (b) The vertically integrated flame chemiluminescence at different $\YHfirst$ (top) along with the flame centre of mass and characteristic width (bottom).}
    \label{fig:combined_figures_1st_var}
\end{figure}

The vertically integrated flame chemiluminescence for this test case is shown in figure~\ref{fig:combined_figures_1st_var}b) and it can be seen that the shape does not change significantly between different $\YHfirst$. Furthermore, the flame center of mass and characteristic width show no clear trend. This observation is aligned again with the resulting DTD parameters from the FTF fittings. The mean time delay $\tau$ remains constant and its spread $\sigma$ does not vary as significantly as in the previous case where $\YHsecond$ is varied.

\section{Conclusions and outlook}

This study experimentally quantifies the sequential flame transfer function (FTF) of an atmospheric sequential combustor. The results reveal that the sequential FTF is sensitive to hydrogen blending in the second-stage fuel line but remains largely unaffected by hydrogen blending in the first-stage fuel line. By employing the classical $n-\tau-\sigma$ model, the sequential FTF can be effectively fitted with two distinct pulses. Higher hydrogen blending in the second-stage fuel mixture results in a more compact and shorter sequential flame. The observed trends in the time delay parameters $(\tau_1,\tau_2)$ and the characteristic spread parameters $(\sigma_1,\sigma_2)$ are in close agreement with the observed changes in the flame length and distribution characteristics. 

This first measurement of CPSC's second-stage FTF paves the way for future studies, such as the FTF measurement of the first-stage flame for stability analyses of the whole thermoacoustic system or the measurement of the second-stage flame describing function (FDF) by forcing it at larger amplitude. Furthermore, by employing the same methodology, the effects of nanosecond repetitively pulsed discharges (NRPD) on the sequential FTF can be quantified. This would then provide a quantitative explanation on the stabilizing effect of NRPD on the thermoacoustic of a sequential combustor, which was demonstrated in \cite{DHARMAPUTRA2023}.

\section*{Acknowledgements}
This project has received funding from the European Research Council (ERC) under the European Union’s Horizon 2020 research and innovation program (grant agreement No [820091]).

\bibliographystyle{elsarticle-num}
\bibliography{cas-refs}

\end{document}


\title{Supplementary Materials}
\author{Bayu Dharmaputra, Sergey Shcherbanev, \\ Bruno Schuermans and Nicolas Noiray}
\date{}
\maketitle

\section{Energy Deposition Measurements}

\begin{figure}[H]
    \centering
    \begin{psfrags}
    \psfrag{VoltageVLLLLL}[][]{\color{black} \scriptsize{Voltage~~~}}
    \psfrag{CurrnetALLLLL}[][]{\color{black} \scriptsize{Current~~~}}
    \psfrag{dVdtVnsLLLLL}[][]{\color{black} \scriptsize{dV$\slash$dt~~~}}
    
    \psfrag{0r}[][]{\color{red} \scriptsize{0}} 
    \psfrag{2500r}[][]{\color{red} \scriptsize{2500}}
    \psfrag{5000r}[][]{\color{red} \scriptsize{5000}}
    \psfrag{-2500r}[][]{\color{red} \scriptsize{-2500}}
    \psfrag{-5000r}[][]{\color{red} \scriptsize{-5000}}  
    \psfrag{dVdtVns}[][]{\color{red} \scriptsize{dV$\slash$dt (V$\slash$ns)}}    

    \psfrag{0}[][]{\color{black} \scriptsize{0}} 
    \psfrag{4000}[][]{\color{black} \scriptsize{4000}}
    \psfrag{8000}[][]{\color{black} \scriptsize{8000}}
    \psfrag{-4000}[][]{\color{black} \scriptsize{-4000}}
    \psfrag{-8000}[][]{\color{black} \scriptsize{-8000}}  
    \psfrag{VoltageV}[][]{\color{red} \scriptsize{Voltage ($V$)}}

    \psfrag{0b}[][]{\color{blue} \scriptsize{0}} 
    \psfrag{0.5b}[][]{\color{blue} \scriptsize{0.5}}
    \psfrag{1b}[][]{\color{blue} \scriptsize{1}}
    \psfrag{-0.5b}[][]{\color{blue} \scriptsize{-0.5}}
    \psfrag{-1b}[][]{\color{blue} \scriptsize{-1}}  
    \psfrag{CurrnetA}[][]{\color{blue} \scriptsize{Current ($A$)}}

    \psfrag{TimeNs}[][]{\color{black} \scriptsize{$t$ (ns)}}
    \psfrag{20}[][]{\color{black} \scriptsize{20}}
    \psfrag{40}[][]{\color{black} \scriptsize{40}}
    \psfrag{60}[][]{\color{black} \scriptsize{60}}
    \psfrag{80}[][]{\color{black} \scriptsize{80}}  
    \psfrag{100}[][]{\color{black} \scriptsize{100}}  
    
    \includegraphics[trim=0cm 0cm 0cm 0cm,clip,width=0.75\textwidth]
    {figures_review/displacement_current.eps}
    \end{psfrags}
    \caption{Measured voltage, current, and the time derivative of the voltage without plasma initation.}
    \label{fig:Displacement}
\end{figure}

Following the procedure in~\cite{rusterholtz2013ultrafast}, the displacement current was acquired by progressively lowering the generator voltage until it fell below the breakdown voltage threshold. As the displacement current is proportional to the derivative of the voltage waveform, an instrumental time lag between the measured current waveform and the time derivative of the voltage signal was calibrated, with a time adjustment of 2.5 ns. This adjustment corresponds to the time lag between current and voltage traces

Characteristic voltage and current waveforms without plasma initiation are shown in figure~\ref{fig:Displacement}, demonstrating the synchronization process. The absolute value of the conduction current during the discharge was found to be significantly higher than the values of the displacement current, and thus the value of the latter was neglected while calculating the deposited energy values.

\section{Acoustic Modal Analyses}

\begin{figure}[H]
    \centering
    \begin{psfrags}

    \psfrag{NOemission}[b][]{\hspace{-0.6cm}\scriptsize{NO~($\mathrm{ppmvd}$)}}
    \psfrag{COMmm}[][]{\scriptsize{COM (mm)}}
    
    \psfrag{WithPlasma}[][]{\scriptsize{with plasma}}
    \psfrag{WithoutPlasmaxxx}[][]{\hspace{-0.4cm}\scriptsize{without plasma}}    
    \psfrag{0}[][]{\scriptsize{0}~~}   
    \psfrag{0.6}[][]{\scriptsize{0.5}} 
    \psfrag{1}[][]{\scriptsize{1}}  
    \psfrag{0.75}[][]{\scriptsize{0.75}~~}
    \psfrag{3}[][]{\scriptsize{3}}
    \psfrag{-0.15}[][]{\scriptsize{-0.15}~~}
    \psfrag{-0.4}[][]{\scriptsize{-0.4}~~}
    \psfrag{-3}[][]{\scriptsize{-3}~~}

    \psfrag{0.15}[][]{\scriptsize{0.15}~~}
    \psfrag{1.5}[][]{\scriptsize{1.5}~}    
    \psfrag{-1.5}[][]{\scriptsize{-1.5}~}    
    \psfrag{0.5}[][]{\scriptsize{0.5}}
    \psfrag{Phat}[b][]{\scriptsize{$\hat{P}$} [-]}
    \psfrag{Phatabs}[b][]{\scriptsize{$|\hat{P}|$} [-]}
    \psfrag{Temperature}[b][]{\scriptsize{$T$ (K)}}
    
    \psfrag{pi}[][]{\scriptsize{$\pi$}}    
    \psfrag{pi2}[][]{\scriptsize{$\frac{\pi}{2}$}}
    \psfrag{f231Hz}[][]{\scriptsize{$f$ = 231 Hz}}
    \psfrag{f327Hz}[][]{\scriptsize{$f$ = 327 Hz}}
    \psfrag{Zposition}[][]{\scriptsize{$z$ (m)}}
    \psfrag{xposition}[][]{\scriptsize{$x$ (m)}}

    \psfrag{300}[][]{\scriptsize{300}}
    \psfrag{800}[][]{\scriptsize{800}}
    \psfrag{1300}[][]{\scriptsize{1300}}
    \psfrag{1800}[][]{\scriptsize{1800}}
    \psfrag{2300}[][]{\scriptsize{2300}}
    
    \includegraphics[trim=0cm 0cm 0cm 0cm,clip,width=1\textwidth]
    {figures_review/domain_temperature_add.eps}
    \end{psfrags}
    \caption{Side view of the computational domain for the pure acoustic analysis (top). The imposed temperature distribution along the longitudinal direction (bottom).}
    \label{fig:domain_temp}
\end{figure}

To understand the characteristics of the two observed modes at around 260~Hz and 330~Hz, modal analyses by solving the acoustic helmholtz equation were performed using COMSOL. Figure \ref{fig:domain_temp} shows the side view of the 3D computational domain and the imposed temperature distribution along the longitudinal direction. The temperature was obtained by computing the adiabatic flame temperature with Cantera. The temperature inside the mixing channel is obtained by assuming a homogeneous mixing between the hot products from the first-stage combustor and the dilution air.

The purpose of the analysis is not to exactly model the complete acoustics of the system. This would require an appropriate description of the complex-valued frequency dependent boundary conditions and loss mechanisms at the area changes as well as the transfer functions of both flames. The intend is to provide a visual rendering of length scales of possible modes involved and that it is plausible that the two observed peaks are of acoustic origin, despite being rather closely spaced in frequency. The inclusion of aeroacoustic and thermoacoustic gains and losses, as demonstrated in the work by \cite{Meindl_aerothermo, EMMERT2017_MODES}, can lead to alterations in the system eigenmodes.

\begin{figure}[H]
    \centering
    \begin{psfrags}

    \psfrag{NOemission}[b][]{\hspace{-0.6cm}\scriptsize{NO~($\mathrm{ppmvd}$)}}
    \psfrag{COMmm}[][]{\scriptsize{COM (mm)}}
    
    \psfrag{WithPlasma}[][]{\scriptsize{with plasma}}
    \psfrag{WithoutPlasmaxxx}[][]{\hspace{-0.4cm}\scriptsize{without plasma}}    
    \psfrag{0}[][]{\scriptsize{0}~~}   
    \psfrag{0.6}[][]{\scriptsize{0.5}} 
    \psfrag{1}[][]{\scriptsize{1}}  
    \psfrag{0.75}[][]{\scriptsize{0.75}~~}
    \psfrag{3}[][]{\scriptsize{3}}
    \psfrag{-0.15}[][]{\scriptsize{-0.15}~~}
    \psfrag{-0.4}[][]{\scriptsize{-0.4}~~}
    \psfrag{-3}[][]{\scriptsize{-3}~~}

    \psfrag{0.15}[][]{\scriptsize{0.15}~~}
    \psfrag{1.5}[][]{\scriptsize{1.5}~}    
    \psfrag{-1.5}[][]{\scriptsize{-1.5}~}    
    \psfrag{0.5}[][]{\scriptsize{0.5}}
    \psfrag{Phat}[b][]{\scriptsize{$\hat{P}$} [-]}
    \psfrag{Phatabs}[b][]{\scriptsize{$|\hat{P}|$} [-]}
    \psfrag{Angle}[b][]{\scriptsize{$\angle \hat{P}$}}

    \psfrag{pi}[][]{\scriptsize{$\pi$}}    
    \psfrag{pi2}[][]{\scriptsize{$\frac{\pi}{2}$}}
    \psfrag{f231Hz}[][]{\scriptsize{$f$ = 231 Hz}}
    \psfrag{f327Hz}[][]{\scriptsize{$f$ = 327 Hz}}
    \psfrag{Zposition}[][]{\scriptsize{$z$ (m)}}
    \psfrag{xposition}[][]{\scriptsize{$x$ (m)}}

    \includegraphics[trim=0cm 0cm 0cm 0cm,clip,width=1\textwidth]
    {figures_review/eigenmodes.eps}
    \end{psfrags}
    \caption{Computed longitudinal eigenmodes (top), longitudinal profile of the absolute values of the eigenmodes (middle), and the angle of the eigenmodes (bottom). The circles in the middle and bottom plot correspond to the experimentally measured signals. The pressure mode is normalized with respect to its value at the first microphone location (z = 0.75~m). The left and right plots show the eigenmodes at the eigenfrequency of 231 Hz and 327 Hz respectively.}
    \label{fig:eigenmodes}
\end{figure}

The disparities observed between the solutions from the Helmholtz solver and the experimental results can be attributed to the omission of aeroacoustic and thermoacoustic gains in the model. These disparities have been observed in the literature such as \cite{XIONG_thermoacoustic}, and \cite{SCHULZ_Thermoacoustic}. In summary, the intensity of the peaks at distinct locations is influenced by the combined effects of the pure acoustic modes, aeroacoustic and thermoacoustic gains and losses that constitute the system's eigenmodes.

\bibliographystyle{elsarticle-num}
\bibliography{cas-refs}